\documentclass[preprintnumbers,floatfix,showkeys,twocolumn]{revtex4}

\usepackage{graphics}
\usepackage{epsfig}
\usepackage{amsmath}
\usepackage{amssymb}
\usepackage{dcolumn}
\usepackage{graphicx,textcomp}
\usepackage{bm}
\usepackage{color}
\usepackage{multirow}
\usepackage{revsymb}
\usepackage[T1]{fontenc}

\begin{document}

\title{Theoretical investigation on the optical absorption spectra in cyclo[$n$]carbons ($n$=10, 14, 18)}

\author{Xuhai Hong\footnote{Corresponding author.\\
E-mail address: hongxuhai@163.com}, Lang Su, Jie Li}
\affiliation{School of Physics and Electronic Technology, Liaoning Normal University, Dalian 116029, China}

\begin{abstract}
The optical absorption spectra of cyclo[$n$]carbons ($n$=10, 14, 18) are investigated in the framework of time-dependent density functional theory. The collective plasmon excitations well develop as the increases of the ring size and the symmetry group of cyclo[$n$]carbons. An increase in intensity for the main peaks with the growing number of atoms in cyclo[$n$]carbons is observed. With the increase of the radius of the monocyclic ring, as more electrons participate in the dipole oscillation the main excitation peaks are red-shifted to the lower energy. The highly symmetrical structures of cyclo[$n$]carbons (D$_{nh}$) possess degenerate levels, leading to simpler spectra with fewer peaks. The Fourier transform of the induced electron density of the cyclo[$n$]carbons ($n$=10, 14, 18) is investigated at the excitation frequencies.
\end{abstract}


\maketitle

\section{Introduction}
For half a century, the nanoscale carbon clusters have attracted remarkable attention in the fields of spectroscopy, theoretical chemistry, and synthetic chemistry due to their potential applications and possible astrophysical significance\cite{NF21}. Some studies indicate that the carbon clusters in the interstellar medium form through the chemical evolution of polycyclic aromatic hydrocarbon molecules, including dehydrogenation from ultraviolet photolysis and subsequent graphene fragmentation\cite{AGG13}. Synthetic chemists have strived to create nanoscale carbon allotropes with outstanding electronic and optical properties for prospective applications. The successful synthesis of carbon allotropes like fullerene, carbon nanotube, and graphene presents challenges for researchers exploring uncharted territories. The number of isomers is widely recognized to significantly increase with the growing atom number $n$ in a system. The monocyclic clusters, composed solely of carbon atoms within a range from 10 to 22, long believed to exist in the universe\cite{HRH23}. As the carbon atoms in pure clusters increase, the polycyclic structures dominate for tens of atoms, while the cage structures prevail when $n$ exceeds 60. The evidence suggests that the carbon cage structures form through nucleation and coalescence among medium-sized carbon clusters, rather than sequential addition of small fragments like C$_{2}$ and C$_{3}$\cite{YT00}. Increasing evidence implies that monocyclic carbon clusters play a crucial role in the early stages of the mechanism leading to the formation of carbon cage structures\cite{YT00,GvH91}.

The monocyclic carbon clusters, systematically named as the cyclo[$n$]carbons, where $n$ defines the number of carbon atoms forming the monocyclic structure, are predicted to exhibit distinctive aromatic stabilization for cyclo[$n$]carbons ($n$=10, 14, 18). Due to the reduction in angle strain, the carbon clusters smaller than C$_{10}$ are likely to exist as low-energy linear structures\cite{GJ19}. The aromaticity arises from the presence of two sets of mutually perpendicular $\pi$-orbitals, one in-plane and one out-of-plane. Pioneering work involved gas-phase synthesis of cyclo[18]carbon through the stepwise elimination of three anthracenes from the precursor hexadehydro[l8]annulene in a retro-Diels-Alder reaction under flash vacuum pyrolysis condition\cite{FD89}. The subsequent attempts to synthesize cyclo[$n$]carbons using the novel compounds as precursors lacked the definite evidence for the direct formation\cite{YR90,FD94,YT00,YT03}. The ground state structure of cyclo[18]carbon has been a subject of theoretical controversy, with predicted structures divided into the polyynic ring with the alternating single and triple bonds (D$_{9h}$ \cite{FD89,TT00,DAP95,VP91,MF92} and C$_{9h}$\cite{DAP95,SA08} symmetry groups) and the cumulenic ring with the consecutive double bonds (D$_{18h}$ \cite{FD89,DAP95,VP91,MF92,SA08,JH94,AEB05,EJ00,ROJ99} and D$_{9h}$\cite{AEB05,SS06,JPM06,PWF09} symmetry groups). Recently, the cyclo[18]carbon was successfully synthesized through atom manipulation, eliminating carbon monoxide from a cyclocarbon oxide molecule C$_{24}$O$_{5}$ on a bilayer NaCl surface on Cu(111) at low temperature (5 K)\cite{KK19}. The high-resolution atomic force microscopy revealed its polyynic ring structure of D$_{9h}$ with bond length alternation, also supported by their computational simulation results.

Following the unprecedented experiment, numerous theoretical studies about cyclo[18]carbon were conducted. Baryshnikov et al. computationally studied its electronic structure, aromaticity, and absorption properties on a NaCl surface at density functional theory and ab initio levels\cite{GVB20}. Hussain et al. explored the tip-enhanced Raman spectroscopy of cyclo[18]carbon based on tip distance and lateral positions of a silver cluster above the ring plane using density functional theory\cite{SH20}. Pereira et al. explained the stability of the polyynic structure through the symmetry breaking driven by the second-order Jahn-Teller effect\cite{ZSP20}. Br\'{e}mond et al. developed the range-separated nonempirical schemes to address structural differences independent of the exchange-correlation functional form\cite{EB19}. Shi et al. studied the electronic structure and excitation characteristics of cyclo[18]carbon\cite{BS20}. Fedik et al. theoretically reported the poly[$n$]catenanes constructed from cyclo[18]carbon, promising the candidates for supramolecular synthesis\cite{NF20}. Stasyuk et al. suggested cyclo[18]carbon as the smallest all-carbon electron acceptor due to its strong electron acceptor properties\cite{AJS20}. Li et al. highlighted the higher stability and aromaticity in cyclo[$n$]carbons ($n$=10 and 14) compared to the acknowledged cyclo[18]carbons, positioning them as the potential molecular semiconductor devices\cite{MYL20}. Liu et al. studied the electronic spectrum and optical nonlinearity of cyclo[18]carbon by the means of (time-dependent) density functional theory calculations\cite{ZY20}. Despite the significant results from these theoretical studies, the research on optical absorption properties remains at an early stage.

Due to the challenge of confirming structural information through small energy differences between two nanostructures using theoretical calculations, alternative properties must be considered. The optical absorption spectra, serving as the unique fingerprints of nanostructures, offer a robust tool for identifying structurally similar ones by comparing the measured and theoretical spectra. The transitions between electron states intrinsically determine the optical absorption properties, characterized by sharp peaks, making them both fundamentally and technologically significant\cite{XG00}. Providing the theoretical spectroscopic support for determining the structures of cyclo[$n$]carbons ($n$=10, 14, 18) is crucial for the synthesis experiments.

This study aims to numerically investigate the optical absorption properties of cyclo[$n$]carbons ($n$=10, 14, 18) within the framework of time-dependent density functional theory (TDDFT). The collective plasmon excitations are explored with increasing ring size and different isomers of cyclo[$n$]carbons. The research demonstrates that the optical absorption properties of nanostructures serve as an indirect approach to identify the corresponding microstructures. The collective electron behavior within specific frequency ranges in nanostructures is expected to have potential applications.

\section{Theory}

To investigate the nature of optical absorption and the size-dependent trend of cyclo[$n$]carbons ($n$=10, 14, 18), the calculations are performed using the Octopus program\cite{MA03,AC06}, a real-time and real-space code effectively implementing static DFT and time-dependent DFT. The calculations begin with the geometry optimization using full-fledged DFT. The total energy is taken as an objective function of the convergence criterion to minimize during the geometry optimization. In order to obtain the equilibrium structures, the coordinates of carbon atoms in the present optimization are substituted by those obtained from the last optimization until the two coordinates are very close. Subsequently, the Kohn-Sham equations initially disturbed by a delta electric field, are evolved on the real-space grids using TDDFT to obtain the spatially averaged absorption spectra. TDDFT, a reliable tool for exploiting electron excitation in the linear domain, implements a complicated time-dependent many-electron process under the ingenious Kohn-Sham effective potential, balancing calculation accuracy and efficiency (Ref. \cite{Runge} for details).

The optical absorption spectra of cyclo[$n$]carbons at zero temperature are calculated using TDDFT. The system is excited by applying a weak delta electric field $E(t)=E_{0}\gamma\delta(t)$, where $\gamma=x, y ,z$ determines the polarization direction, and $E_{0}$ is the amplitude of the electric field, small enough for the perturbation within the linear region. This weak perturbation leads each Kohn-Sham orbital of the system initially in the ground state to undergo an instantaneous phase-shift, equivalent to the excitation by all frequencies with equal weight
\begin{equation}
\psi_{i}\left(\textbf{r},t=0^{+}\right)=
{\rm e}^{iE_{0}\gamma}\psi_{i}\left(\textbf{r},t=0^{-}\right).\\
\end{equation}

The Kohn-Sham obitals begin to propagate over a finite period $T$ when the active electrons participate in collective resonance, moving back and forth uniformly under the background of positive ionic cores. Throughout the evolution, the time-dependent density $n(\textbf{r}, t)$ is recorded readily
\begin{equation}
n(\textbf{r},t)=\sum^{N}_{i}\left\vert\psi_{i}(\textbf{r},t)\right\vert^{2}.
\end{equation}

The dynamical dipole polarizability is calculated by
\begin{equation}
\alpha_{\gamma}(\omega)=-\frac{1}{E_{0}}\int^{T}_{0} {\rm d}^{3}r\gamma\delta n(\textbf{r},\omega),
\end{equation}
where $\delta n(\textbf{r},\omega)$ is the Fourier transform of the difference of time-dependent electron density and ground-state electron density of the system
\begin{equation}\label{eq:4}
\delta n(\textbf{r},\omega)=\int^{T}_{0}{\rm d}t {\rm e}^{i\omega t}f(t)\delta n(\textbf{r},t),
\end{equation}
where $f(t)$ is a third-order polynomial damping function, which reasonably considers the physical broadening of excitation peaks.

The optical absorption spectra is obtained by averaging over the three spatial coordinates
\begin{equation}
\sigma(\omega)=\frac{4\pi\omega}{c}\frac{1}{3}
{\rm Im}\left(\sum_{\gamma}\alpha_{\gamma}(\omega)\right),
\end{equation}
where $c$ is the speed of light.

The simulation parameters are given below. The norm-conserving pseudopotentials are used to describe the interaction between active electrons and carbon ionic cores\cite{NT91}. The simulation box, discretized by a uniform mesh grid of 0.4 a.u., centers on the geometric center of the isolated cyclo[$n$]carbons ($n$=10, 14, 18), with the monocyclic cluster's plane at $z$=0 a.u. Kohn-Sham equations evolve for a total time $T$ of 1000 a.u., with a time step of 0.05 a.u. These parameters are verified for the convergence of calculations to ensure stable time evolution and high-resolution spectra. The Perdew-Zunger exchange-correlation functional\cite{JP81} within the local-density approximation (LDA)\cite{EKUG94} describes the exchange and correlation (xc) potential both in the static and time-dependent calculations.

\section{Results and discussion}

The present optimized equilibrium structures of cyclo[$n$]carbons ($n$=10, 14, 18) are compared with the previous theoretical calculations (see Table 1). Each monocyclic carbon cluster yields two isomers, corresponding to the symmetry groups D$_{0.5nh}$ and D$_{nh}$ ($n$=10, 14, 18). The D$_{nh}$ structures have the identical bond lengths and angles. In the numerical calculations, the direction of optimization results may be algorithm-sensitive for similar isomers with tiny total energy difference. For carbon clusters, the smallest monocyclic structure is C$_{10}$. The carbon cluster C$_{10}$ (D$_{5h}$) is more like a pentagon than a circular ring, because of the relatively large difference between the alternating bond angles (164.6$^{\circ}$, 123.4$^{\circ}$). The present calculated results demonstrate the D$_{0.5nh}$ symmetry groups in these monocyclic carbon clusters are the more stable structures. The relative energies for different symmetry groups are given in Table 1. The optimization results are in good agreement with the previous studies.

\begin{table}
\caption{The comparison of the present optimized and other calculated structure parameters for cyclo[$n$]carbons ($n$=10, 14, 18). The bond lengths (a.u.), the angles (degree) and relative energies ($\Delta$E in eV) for different symmetry groups are given.}
\begin{tabular}{lcccc}
\hline
 Structure & Bond length & Angle & $\Delta$E & Reference\\
\hline
C$_{18}$(D$_{9h}$)& 2.312, 2.541 & 160$^{\circ}$ & 0.00 & Present\\
                  & 2.266, 2.574 & 160$^{\circ}$ &   -  & Ref. \cite{FD89}\\
                  & 2.344, 2.514 & 160$^{\circ}$ &   -  & Ref. \cite{TT00}\\
                  & 2.264, 2.608 & 160$^{\circ}$ &   -  & Ref. \cite{DAP95}\\
                  & 2.257, 2.599 & 160$^{\circ}$ &   -  & Ref. \cite{VP91}\\
                  & 2.268, 2.570 & 160$^{\circ}$ &   -  & Ref. \cite{MF92}\\
C$_{18}$(D$_{18h}$)& 2.411 & 160$^{\circ}$ & 1.65 & Present\\
                   & 2.391 & 160$^{\circ}$ &  -   & Ref. \cite{FD89}\\
                   & 2.402 & 160$^{\circ}$ &  -   & Ref. \cite{DAP95}\\
                   & 2.413 & 160$^{\circ}$ &  -   & Ref. \cite{VP91}\\
                   & 2.400 & 160$^{\circ}$ &  -   & Ref. \cite{MF92}\\
                   & 2.429 & 160$^{\circ}$ &  -   & Ref. \cite{SA08}\\
                   & 2.432 & 160$^{\circ}$ &  -   & Ref. \cite{JH94}\\
                   & 2.429 & 160$^{\circ}$ &  -   & Ref. \cite{AEB05}\\
                   & 2.400 & 160$^{\circ}$ &  -   & Ref. \cite{EJ00}\\
                   & 2.419 & 160$^{\circ}$ &  -   & Ref. \cite{ROJ99}\\
C$_{14}$(D$_{7h}$)& 2.338, 2.505 & 154.3$^{\circ}$ & 0.00 & Present\\
                  & 2.349, 2.519 & 154.3$^{\circ}$ &  -   & Ref. \cite{TT00}\\
                  & 2.270, 2.604 & 154.3$^{\circ}$ &  -   & Ref. \cite{DAP95}\\
C$_{14}$(D$_{14h}$)& 2.421 & 154.3$^{\circ}$ & 0.48 & Present\\
                   & 2.404 & 154.3$^{\circ}$ &  -   & Ref. \cite{DAP95}\\
                   & 2.432 & 154.3$^{\circ}$ &  -   & Ref. \cite{SA08}\\
                   & 2.436 & 154.3$^{\circ}$ &  -   & Ref. \cite{JH94}\\
                   & 2.400 & 154.3$^{\circ}$ &  -   & Ref. \cite{EJ00}\\
                   & 2.434 & 154.3$^{\circ}$ &  -   & Ref. \cite{SS06}\\
C$_{10}$(D$_{5h}$) & 2.434 & 123.4$^{\circ}$, 164.6$^{\circ}$ & 0.00 & Present \\
                   & 2.453 & 127.7$^{\circ}$, 160.3$^{\circ}$ &  -   & Ref. \cite{TT00} \\
                   & 2.438 & 119.4$^{\circ}$, 168.6$^{\circ}$ &  -   & Ref. \cite{DAP95}\\
                   & 2.449 & 129.52$^{\circ}$, 158.48$^{\circ}$ &  -   & Ref. \cite{SA08}\\
                   & 2.451 & 130.1$^{\circ}$, 157.9$^{\circ}$ &  -   & Ref. \cite{JH94}\\
                   & 2.410 & 130.8$^{\circ}$, 157.2$^{\circ}$ &  -   & Ref. \cite{EJ00}\\
                   & 2.440 & 126.2$^{\circ}$, 161.8$^{\circ}$ &  -   & Ref. \cite{SS06}\\
C$_{10}$(D$_{10h}$)& 2.420 & 144$^{\circ}$ & 0.47 & Present\\
                   & 2.412 & 144$^{\circ}$ &  -   & Ref. \cite{DAP95}\\
                   & 2.441 & 144$^{\circ}$ &  -   & Ref. \cite{SA08}\\
                   & 2.410 & 144$^{\circ}$ &  -   & Ref. \cite{EJ00}\\
                   & 2.440 & 144$^{\circ}$ &  -   & Ref. \cite{SS06}\\
\hline
\end{tabular}
\end{table}

\begin{figure}[htbp]
\begin{center}
\includegraphics{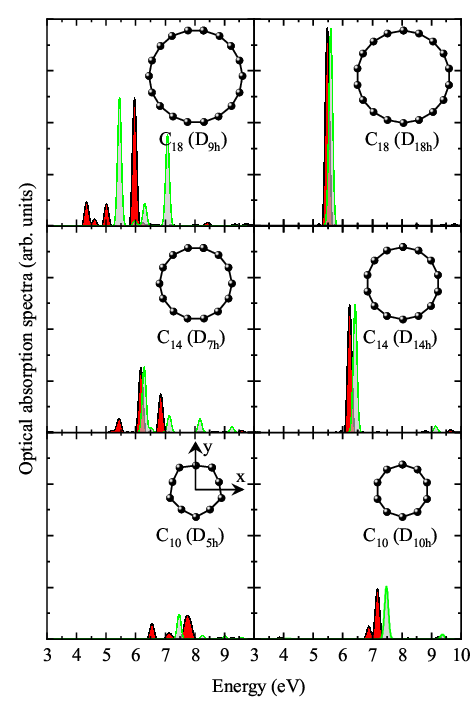}
\caption{The optical absorption spectra of cyclo[$n$]carbons ($n$=10, 14, 18). The collective plasmon excitation peaks in the $x$- and $y$-directions are coincident each other. The results calculated using LDA and M06-2X xc functional are indicated by the black/green solid lines with the under area filled with red/gray, respectively. The collective plasmon excitation peaks in the $z$-direction are not shown due to the much weaker intensity of its peaks compared to the other two directions. The inset is the schematic diagram of the monocyclic carbon clusters. (colour online)}
\end{center}
\end{figure}

The experimental spectra for the recently synthesized cyclo[18]carbon face challenges, including the low efficiency and the high difficulty in preparing one cluster at a time, and the specific substrate used not meeting the gas phase conditions required for measuring optical absorption spectra. Based on the above reasons, it is necessary to theoretically calculate the optical absorption spectra with different levels of xc functional to verify each other. Fig. 1 displays the calculated optical absorption spectra of cyclo[$n$]carbons ($n$=10, 14, 18) using two kinds of xc functionals. The TDDFT calculations using the LDA xc functional demonstrate relative reliability for the low-energy optical absorption spectra, with an estimated error of 0.2-0.4 eV, since the non-local nature of xc potential is simplified in the LDA treatment\cite{XH11}. For comparison, the calculations using more accurate M06-2X xc functional within hybrid meta generalized gradient approximation in conjunction with def2-TZVP basis set are performed using Gaussian 16 (B.01) program package\cite{MJ16}. The symmetrical placement about the $z$-direction (perpendicular to the ring plane) in the real-space simulation box results in completely overlapping spectra along the $x$- and $y$-directions (parallel to the ring plane), while only the higher-energy peaks with much weaker intensity along the $z$-direction are shown due to the monocyclic carbon clusters' geometry. For simplicity, only the optical absorption spectra along the $x$-direction are displayed in Fig. 1. The isolated peaks shown in Fig. 1 corresponds to the energy level differences through dipole transition. For the cyclo[$n$]carbons (D$_{nh}$), the results calculated using LDA and M06-2X xc functional agree well except for an tiny energy shift of about 0.2 eV. In the left panels (D$_{0.5nh}$), the number of excitation peaks is consistent for the two xc functionals, but the energy range undergoes an energy shift. The following analyses are based on the results of LDA calculations.

\begin{figure}[htbp]
\begin{center}
\includegraphics{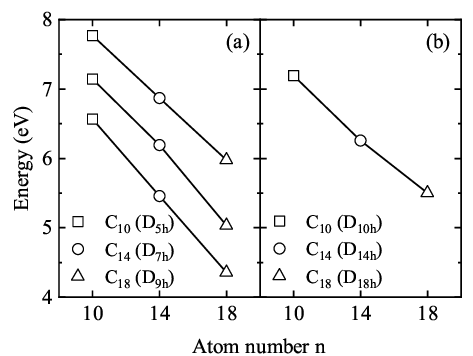}
\caption{The energy evolution of the main optical absorption peaks with the increase of the atom number in cyclo[$n$]carbons ($n$=10, 14, 18) for the symmetry groups of D$_{0.5nh}$ (a) and D$_{nh}$ (b).}
\end{center}
\end{figure}

The optical absorption spectra of nanostructures, interpreted as localized surface plasmon resonance, are highly sensitive to the variations in nanostructure geometry. The geometric extensions of the electron density along three spatial directions correlate closely with the peak positions in these spectra. The optical absorption spectra in the $z$-direction, generated by the narrow extension (consisting of just one layer of atoms) perpendicular to the monocycle plane, reside at the higher energy region. In the other two directions, the optical absorption spectra are situated at the lower energy region. Due to their structural characteristics, the monocyclic carbon clusters exhibit the tunable plasmonic resonance by modifying system details such as the symmetric group and the ring radius size. An increase in intensity for the main peaks in both $x$- and $y$-directions with the growing number of atoms in cyclo[$n$]carbons is observed. This evolution behavior of peaks arises from the accumulation of collective plasmon excitation as more valence electrons participate in the collective plasmon oscillation\cite{JY07}. The integral areas below the spectral curves in Fig. 1 obey the dipole sum rule for the total number of the active electrons involved in collective motion\cite{AC05}. Moreover, as the radius of monocyclic ring increases, there is an overall shift towards the lower energy regions for these main peaks. Fig. 2 illustrates the energy evolution of the main peaks with the increase of atom number $n$ in the cyclo[$n$]carbons. These red-shift evolution behaviors can be interpreted as the quantum confinement effect intrinsic to nanoscale regimes. When the electron density distribution extends further along a specific direction, the dipole polarizability is enhanced, and the absorption peak positions experience red shifts. This effect is particularly pronounced for the linear or planar clusters or molecules due to their spatial extensions. Taking the organic molecule as an example, it is demonstrated that the optical property of the typical polycyclic aromatic hydrocarbons (PAHs) small molecular systems well evolve along with the molecular size\cite{AJ16,YN23}. Besides, the energy and the intensity of molecular plasmon-like excitations in PAH derivatives could be largely tuned through molecular size and heteroatom-modification\cite{YN23}. In contrast, the isotropic clusters like fullerene C$_{60}$ exhibit completely overlapping optical absorption spectra across all three spatial directions\cite{AC05}. Despite the identical atom number and the equivalent radius, the optical absorption spectra of cyclo[$n$]carbons (D$_{nh}$) exhibit fewer peaks compared to those of cyclo[$n$]carbons (D$_{0.5nh}$). This can be attributed to the highly symmetrical structures of cyclo[$n$]carbons (D$_{nh}$), possessing degenerate levels leading to simpler spectra with fewer peaks\cite{XH20}. It is further verified from the above discussion, the optical properties can be predominantly modulated by the size and geometry of nanostructures. Meanwhile, it is significantly meaningful for the tunable optical properties of nanostructures to be pushed forward to the potential applications in areas such as photonics devices, photocatalysis, and biological sensing\cite{YS13,YZ14,XY15,LV12}.

\begin{figure}[htbp]
\begin{center}
\includegraphics{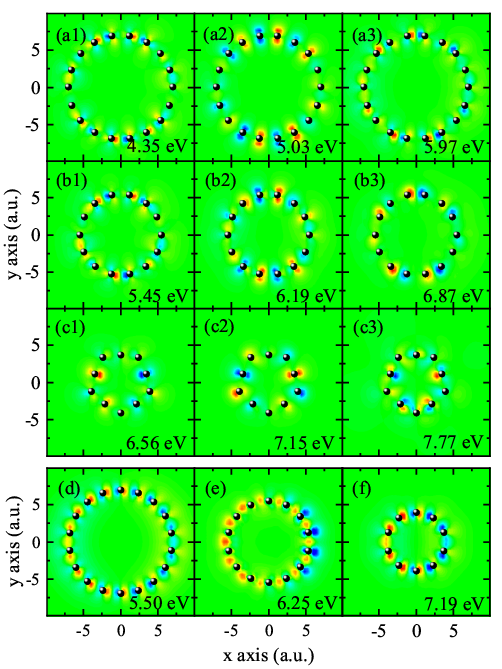}
\caption{The Fourier transform of the induced electron density of cyclo[18]carbons ((a1)-(a3) for D$_{9h}$, (d) for D$_{18h}$), cyclo[14]carbons ((b1)-(b3) for D$_{7h}$, (e) for D$_{14h}$), cyclo[10]carbons ((c1)-(c3) for D$_{5h}$, (f) for D$_{10h}$) at the excitation energies for the main optical absorption peaks. The excitation energies are indicated in each panel. The electron density amplitude increases from blue to red. The black solid spheres indicate carbon atoms. (colour online)}
\end{center}
\end{figure}

Each isolated peak in the aforementioned optical absorption spectra corresponds to a collective electron resonance mode. For a more intuitive visualization of these resonance modes, the Fourier transform of the induced electron density of cyclo[$n$]carbons ($n$=10, 14, 18) is presented in Fig. 3. The observation of induced electron density is of fundamental importance for correctly characterizing the plasmon resonance modes. The excitation energy indicated in each panel is the energy position of the main excitation peak. Calculated at each time step in real-time propagation using Eq. (\ref{eq:4}), the induced electron density oscillates symmetrically with respect to the central axis, demonstrating typical dipole oscillation characteristics. The electronic oscillation modes belongs to particularly frequencies and represents individual peaks in the absorption spectra. The frequency-dependent electron density patterns provide an identification method for clusters and suggest the potential applications for cyclo[n]carbons as the nanodevice candidates.

\section{Summary}
The balanced structures of the cyclo[$n$]carbons ($n$=10, 14, 18) clusters were optimized using DFT, divided into two symmetry groups: D$_{0.5nh}$ and D$_{nh}$. The optical absorption spectra were calculated using TDDFT, revealing that the collective plasmon excitations depend strongly on both the symmetric group and ring radius size. The cyclo[$n$]carbons (D$_{nh}$) exhibit fewer peaks in their optical absorption spectra compared to those of cyclo[$n$]carbons (D$_{0.5nh}$). As the radius of the monocyclic ring increases, the intensity of spectral peaks enhances, and with more electrons participating in dipole oscillation, the main absorption peaks red-shift to the lower energy region. Fourier transform analysis reveals typical dipole oscillation behavior in the induced electron density of cyclo[$n$]carbons at excitation frequencies, indicating promising applications for these collective electron behaviors in specific frequency regions.

\section*{Acknowledgments}
This work was supported by the National Natural Science Foundation of China [grant numbers 11804136, 11774344].

\end{document}